\documentclass[12pt,a4paper]{article}
\usepackage{a4wide}
\usepackage{amssymb}
\usepackage{amsmath} 
\usepackage{graphicx}
\usepackage[breaklinks]{hyperref}
\usepackage{graphicx,caption,xcolor}
\usepackage[style=nature, citestyle=numeric-comp, sorting=none, backend=biber, maxnames = 5]{biblatex}
\usepackage{xurl}
\begin{document}
	{\renewcommand{\thefootnote}{\fnsymbol{footnote}}
		\begin{center}
			{\LARGE Collective field theory of gauged multi-matrix models:\\[2mm] Integrating out off-diagonal strings}\\
			\vspace{1.5em}
			Suddhasattwa Brahma$^1$\footnote{e-mail address: {\tt suddhasattwa.brahma@gmail.com}},
			Robert Brandenberger$^2$\footnote{e-mail address: {\tt rhb@hep.physics.mcgill.ca}},
			Keshav Dasgupta$^2$\footnote{e-mail address: {\tt keshav@hep.physics.mcgill.ca}},\\
			Yue Lei$^2$\footnote{e-mail address: {\tt yue.lei@mail.mcgill.ca}}
			and Julia Pasiecznik$^2$\footnote{e-mail address: {\tt julia.Pasiecznik@mail.mcgill.ca}}
			\\
			\vspace{0.5em}
			$^1$Higgs Centre for Theoretical Physics, School of Physics \& Astronomy,\\
			University of Edinburgh, Edinburgh EH9 3FD, Scotland, UK\\
			\vspace{0.5em}
			$^2$Department of Physics, McGill University, \\
			Montr\'eal, QC, H3A 2T8, Canada
			\vspace{1.5em}
		\end{center}
	}
	
	\setcounter{footnote}{0}
	
	\begin{abstract}
		\noindent We study a two-matrix toy model with a BFSS-like interaction term using the collective field formalism. The main technical simplification is obtained by gauge-fixing first, and integrating out the off-diagonal elements, before changing to the collective field variable. We show that the resulting $(2+1)$-dimensional collective field action has novel features with respect to non-locality, and that we need to add a mass term to get a time-local potential. As is expected, one recovers the single matrix quantum mechanical collective field Hamiltonian in the proper limit.
	\end{abstract}
	
	
	\section{Introduction}
	
One of the most striking aspects of matrix models is that, in the large $N$ limit, they describe an emergent notion of spacetime geometry. There are many known examples of this by now, from the single matrix quantum mechanics \cite{Das:1990kaa} to the so-called BFSS \cite{Banks:1996vh} or IKKT \cite{Ishibashi:1996xs} models. The collective field formalism \cite{Jevicki:1979mb,Jevicki:1980zg,Sakita:1985exh} has been long identified as an efficient way to understand how spatial dimensions emerge from gauge theories. The collective field forms an over-complete basis (and is, thus, a redundant description) except in the strict $N\rightarrow \infty$ limit since it has infinitely many more degrees of freedom compared to the original matrices. However, this multi-local nature of the collective field is precisely how it can be interpreted as a field theory in a higher dimensional spacetime \cite{deMelloKoch:2023ylr}. For the single matrix quantum mechanics, this formalism has been employed to show how it is equivalent to two-dimensional string theory \cite{Das:1990kaa, Klebanov:1991qa}. In this case, the collective field describes the eigenvalue density of the matrix which has a well-defined double-scaling (large $N$) limit to give a continuum description. 

However, generalizing to quantum mechanics of two or more $N \times N$ Hermitian matrices is a highly nontrivial task. The collective field formalism relies on describing the dynamics through gauge-invariant quantities of the field variables. For quantum mechanics of multiple matrices, the invariant `Wilson lines' can be joined together in a way that the gauge invariants are given by the trace over infinitely many possible combinatorics of each of the constituent matrices \cite{Fukuma:1997en}. A useful nomenclature for thinking about this is that the collective field is given by the trace over all possible `words' which are built out of `letters' corresponding to the constituent matrices. However, such `words', labeling the collective variables, are subject to permutations due to the cyclicity of the trace. Solving the corresponding effective Hamiltonian equations for the collective field, or equivalently the Schwinger-Dyson equations of the Wilson loop variables, is not analytically possible in coupled multi-matrix models. One of the reasons for the system to become complex is that in the large $N$ limit, the degrees of freedom increase rapidly and it is impossible to solve the dynamical equations corresponding to the Wilson loops. The well-known exception is, of course, the single matrix model, where the collective field simplifies considerably since the gauge invariants are given by $\operatorname{Tr}(M^n(t))$. Although numerical methods \cite{Jevicki:1982jj, Mathaba:2023non, Rodrigues:2008uh, Cook:2007et} have been developed over the past few decades to find solutions for the collective field corresponding to interacting multi-matrix models \cite{Koch:2021yeb}, it would be more useful to find analytical results for better understanding the emergence of spatial dimensions in these models. Our findings will turn out to be consistent with exact ``bootstrap'' results for similar models \cite{Han:2020bkb}.

In this paper, we employ an alternate approximation scheme to a toy two-matrix model that would, in turn, allow us to write down an analytical form for the effective Hamiltonian/Lagrangian for the collective field. What we consider is a gauged two-matrix model that has an overall $U(N)$ symmetry such that we are interested in physical states that are annihilated by the singlet constraint, as is typical for these modes. To deal with the enormous redundancy in the degrees of freedom of the collective field, our strategy will be to first write down the action for the matrix elements explicitly after fixing the gauge whereby making one of the matrices diagonal (and further ordering the diagonal elements). Then, we will integrate out the off-diagonal matrix elements and find an effective action for the diagonal elements alone. It is at this stage that we shall invoke the collective field formalism to write down an effective Lagrangian corresponding to the system which is now described only by the diagonal elements of the two matrices. To reiterate, we will take care of the off-diagonal strings first by integrating them out before rewriting the quantum mechanics of the $2N$ diagonal elements in terms of the collective field $\phi(x,y,t)$, where each of the spatial dimensions descend from the eigenvalue densities of the two matrices. In this way, we shall automatically confirm the expectation that the two-matrix quantum mechanical model gives rise to a field theory in an ambient $(2+1)$-d spacetime.

Along the way, we shall encounter a few subtleties. Firstly, although a formal integrating out can be done for the off-diagonal elements considering a toy BFSS model, we find that this leads to a time non-local EFT. In this work, since our goal is to emphasize a new approach to apply the collective field formalism by integrating out the off-diagonal elements first, we shall add a small mass term to one of the matrices to obtain a theory which is local in time. However, this will break the correspondence of this two-matrix theory to the BFSS model and will effectively add a cosmological constant term by hand. The emergent non-locality is perhaps expected from other considerations and we shall have more to say about this later. Another subtlety will be regarding the effective theory of the $2N$ diagonal elements, $N$ of which will be fermionic-valued while the other $N$ are bosonic. We will end up defining a collective field corresponding to this quantum theory of $2N$ particles.

The final form of the (effective) collective field Hamiltonian will be such that one recovers the well-known single matrix case when taking the appropriate limit. For the help of the reader not familiar with the details of the collective field theory of matrix quantum mechanics, we review the general collective field formalism in Appendix A and that for the single matrix action in Appendix B. In the next section, we describe our toy model before carrying out the integrating out in Sec-3. In Sec-4, we compute the collective field Hamiltonian for the integrated-out system. We conclude in Sec-5 and also describe what the next steps of this program will be in upcoming papers.


\section{The toy BFSS model}
	
Our starting point is a quantum mechanical action with two bosonic Hermitian matrices $X$ and $Y$ of size $N$, such that they take the form
\begin{eqnarray}
	    S = \frac{1}{2l_{s}}\int d t \operatorname{Tr}\left[\left(D_t X\right)^2 + \left(D_t Y\right)^2-\frac{2}{l_{s}^{4}}[X, Y]^2\right] \ . \label{Action}
\end{eqnarray}
Recalling the BFSS action
\begin{eqnarray}
        S = \frac{1}{2l_{s}}\int d t \operatorname{Tr}\left[\sum_{i=1}^9 \left(D_t X_{i}\right)^2 - \sum_{i\neq j}\frac{2}{l_{s}^{4}}[X_i, X_j]^2 + {\rm fermionic \ terms}\right] \ , \label{BFSS} 
\end{eqnarray}
it is easy to see that our Lagrangian \eqref{Action} can be thought of as a toy model describing the bosonic part of the BFSS action. A $(0+1)$-d $U(N)$ gauge field is introduced by the action of the covariant derivative: $D_t X := \partial_t X + [A, X]$. We shall keep the same physical interpretation of this toy model as understood for the full BFSS model. On identifying the time-coordinate with that of the target-space, the diagonal elements of the two matrices denote the coordinates of an emergent $2$-d space such that they describe the positions of the D$0$-branes whereas the off-diagonal matrix elements are the strings stretched between these D$0$-branes (see e.g. \cite{Taylor:2001vb, Ydri:2017ncg} for reviews). Whether the emergent space is a smooth gravitating spacetime or not, would still need to be explored. We ignore the fermionic degrees of freedom here since we are mainly interested in understanding how the eigenvalue distribution of the bosonic matrices leads to the emergence of spatial directions.

The action \eqref{Action} is invariant under a $U(N)$ symmetry, generated by a (`Gauss law') constraint\footnote{In quantum gravity, one expects the $U(N)$ symmetry to be gauged such that physical states have a zero charge under this isometry \cite{Frenkel:2023aft}, and we are only concerned with the singlet sector of the theory.}, and hence this is known as a gauged matrix model. All the matrix-valued fields transform under the adjoint of the $U(N)$ symmetry. As an aside, note that we would typically like to consider a configuration such that the D$0$-branes are far apart from each other. These are the interesting configurations from the point of view of an emergent cosmology and describe low-energy states that are localized about ``classical'' configurations that minimize the potential. However, as we shall see, such a model does not yield a time-local theory and we will need to add a small mass term to (at least) one of the matrices to get a tractable truncation. 

To take care of gauge-invariance, we first fix the so-called axial gauge, \textit{i.e.,} choose $A(t)=0$. To restrict one to the gauge orbits of this gauge choice, one needs to further choose the parametrisation:
\begin{eqnarray}\label{diag_X}
      X =  \Omega \ X_{\rm gf} \ \Omega^\dagger =: \Omega \ \lambda \ \Omega^\dagger \ ,
\end{eqnarray}
where $\Omega \in U(N)$ is a unitary matrix sometimes called the angular components of $X$. Since $\lambda$ here is a diagonal matrix, this means that we have fixed the gauge by diagonalizing the $X$ matrix. Having made this choice, $Y$ takes the form
\begin{eqnarray}\label{diag_X_basis}
	Y = \Omega \ Y_{\rm gf} \ \Omega^\dagger \ ,
\end{eqnarray}
where the subscript refers to the gauge-fixed version of $Y$. The freedom to diagonalize one of the matrices comes from the realization that fixing $A=0$ implies that the wavefunction is invariant under the $U(N)$ transformations \cite{Fliss:2024don}. In turn, diagonalizing $X$ requires accounting for the (square of the) Vandermonde determinant \cite{Sakita:1985exh}
\begin{eqnarray}
	\Delta(\lambda) = {\prod_{i<j} (\lambda_i - \lambda_j)}
\end{eqnarray}
in the measure of the path integral. Here, $\lambda_{i}$ denotes the elements of the diagonal matrix $\lambda$. We assume that the eigenvalues are distinct and fix the residual gauge freedom by ordering them such that $\lambda_1 < \lambda_2 < \ldots <\lambda_N$.

	
\section{Integrating out the off-diagonal strings}

In the following, let us restrict $N=2$, \textit{i.e.,} we consider both of the matrices to be $2\times 2$ in dimension. This means we are considering a theory of two D$0$-branes in $(2+1)$-d. This makes the explicit computations readily comprehensible without too much algebra; however, our results can be easily generalized to arbitrary $N\times N$ matrices, as we shall explain at the end of this section. Following the previous discussion, we will diagonalize the $X$ matrix and work in that gauge. Then, we shall integrate out the off-diagonal elements of the $Y$ matrix to end with an effective action for the diagonal degrees of freedom alone. 

\subsection{Non-local physics for massless matrices}
After diagonalizing the $X$ matrix, we Wick  rotate and write down the Euclidean action explicitly in terms of the matrix elements. After taking the trace, we get
	\begin{equation}
		S_{E}=\frac{1}{2l_{s}}\int \mathrm{d}\tau\, \left(\dot{\lambda}_{1}^{2}+\dot{\lambda}_{2}^{2}+\dot{\rho}_{1}^{2}+\dot{\rho}_{2}^{2}+2\dot{Y}_{12}^{\ast} \dot{Y}_{12}+\frac{2}{l_{s}^{4}}\left(\lambda_{1}-\lambda_{2}\right)^{2} \lvert Y_{12}\rvert^{2} \right) \,, \label{eq:effective_action}
	\end{equation}
where the \(\lambda_{i}\) and \(\rho_{i}\) represent the diagonal elements of the $X_{\rm gf}$ and $Y_{\rm gf}$ matrices respectively, and $Y_{12}$ represents the off-diagonal element of the $Y_{\rm gf}$ matrix. The explicit form of the potential shows that it is independent of the $\rho$ coordinates, while only being a function of the difference $(\lambda_1 -\lambda_2)$. The former is just a feature of our gauge-fixing choice and obviously we could just as easily have diagonalized $Y$ instead of $X$. Note that since there is a perfect symmetry between $X$ and $Y$ in our starting action \eqref{Action}, this implies choosing a direction which the potential would be independent of. On the other hand, the latter point corroborates the expectation that D$0$-brane dynamics are independent of translations in the emergent spatial directions.

To integrate out $Y_{12}$, note that we do not restrict the range of values for $(\lambda_1-\lambda_2)$, measured in some string units, and need to keep this arbitrary. At this point, we do not need the D$0$-branes to be either well-separated or stacked on top of each other\footnote{One can employ a different (Born-Oppenheimer) approximation when these frequencies are large $(\lambda_1-\lambda_2) \gg 1$ such that the diagonal elements have slow dynamics compared to that of the ``heavy'' off-diagonal elements \cite{Hampapura:2020hfg}. However, we shall not invoke this approximation here to write down the energy eigenstates.}. The action \eqref{eq:effective_action} shows that we can interpret the strings connecting the D-branes as complex harmonic oscillators whose frequency depends on the separation between the  D$0$-branes, $\omega = (\lambda_1-\lambda_2)$. We obtain an effective action for the diagonal degrees of freedom alone by performing the path integral over the off-diagonal strings $Y_{12}$ \cite{Anninos_2020}:
	\begin{equation}\label{eq:effective_action_massless}
		\begin{aligned}
			e^{-S_{\textrm{eff}}\left[\lambda,\rho \right]} &:= \int \mathcal{D} Y_{12}\, e^{-S_{E}\left[\lambda,\rho,Y_{12}\right]}\\
			&=e^{-\frac{1}{2l_{s}}\int \mathrm{d}\tau\, \left(\dot{\lambda}_{1}^{2}+\dot{\lambda}_{2}^{2}+\dot{\rho}_{1}^{2}+\dot{\rho}_{2}^{2}\right)} \int \mathcal{D} Y_{12}\, e^{-\frac{1}{l_{s}}\int \mathrm{d}\tau\, \left( \dot{Y}_{12}^{\ast} \dot{Y}_{12}+\frac{1}{l_{s}^{4}}\left(\lambda_{1}-\lambda_{2}\right)^{2} \lvert Y_{12}\rvert^{2}\right)}\,.
		\end{aligned}
	\end{equation}
The path integral over \(Y_{12}\) is given by a functional determinant
	\begin{equation}
		\begin{aligned}
			Z&=\int \mathcal{D} Y_{12}\, e^{-\frac{1}{l_{s}}\int \mathrm{d}\tau\, \left( \dot{Y}_{12}^{\ast} \dot{Y}_{12}+\frac{1}{l_{s}^{4}}\left(\lambda_{1}-\lambda_{2}\right)^{2} \lvert Y_{12}\rvert^{2}\right)}\\
			&=\int \mathcal{D} Y_{12}\, e^{-\frac{1}{l_{s}}\int\mathrm{d}\tau\, \left(Y_{12}^{\ast}\left(-\frac{\mathrm{d}^{2}}{\mathrm{d}\tau^{2}}+\frac{1}{l_{s}^{4}}\left(\lambda_{1}-\lambda_{2}\right)^{2}\right)Y_{12}\right)}\\
			&=\textrm{const}\times\left[\det \left(\frac{1}{l_{s}}\left(-\frac{\mathrm{d}^{2}}{\mathrm{d}\tau^{2}}+\frac{1}{l_{s}^{4}}\left(\lambda_{1}-\lambda_{2}\right)^{2}\right)\right) \right]^{-1/2}\, .
		\end{aligned}
	\end{equation}
The Euclidean effective action of the diagonal elements of the matrices can then be expressed as 
	\begin{equation}
		S_\textrm{eff}\left[\lambda,\rho\right]=\int \mathrm{d}\tau\, \left(\frac{1}{2l_{s}}\left(\dot{\lambda}_{1}^{2}+\dot{\lambda}_{2}^{2}+\dot{\rho}_{1}^{2}+\dot{\rho}_{2}^{2}\right)\right)-\frac{1}{2}\mathrm{Tr}\left[\log\left(\frac{1}{l_{s}}\left(-\frac{\mathrm{d}^{2}}{\mathrm{d}\tau^{2}}+\frac{1}{l_{s}^{4}}\left(\lambda_{1}-\lambda_{2}\right)^{2}\right)\right)\right]\,,
	\end{equation}
	where we have used the property of the logarithm to rewrite the determinant (which is a product) into a trace. The first term in the parenthesis above simply corresponds to the momenta terms. The second term above can be rewritten as
	\begin{eqnarray}\label{eq:trlog}
		& &\mathrm{Tr}\left[\log\left(\frac{1}{l_{s}} \left(-\frac{\mathrm{d}^{2}}{\mathrm{d}\tau^{2}} +\frac{1}{l_{s}^{4}}\left(\lambda_{1}\left(\tau\right)-\lambda_{2}\left(\tau\right)\right)^{2}\right)\right)\right] \nonumber\\
		&=&\mathrm{Tr}\left[\log\left(-\frac{1}{l_{s}}\frac{\mathrm{d}^{2}}{\mathrm{d}\tau^{2}}\right)\right]+\mathrm{Tr}\left[\log\left(1+\frac{1}{l_{s}^{4}}\left(-\frac{\mathrm{d}^{2}}{\mathrm{d}\tau^{2}}\right)^{-1}\left(\lambda_{1}\left(\tau\right)-\lambda_{2}\left(\tau\right)\right)^{2}\right)\right] \ .
	\end{eqnarray}
	We focus on the second term in the last line above for now. The inverse of the differential operator, which appears in this expression, is given by an integral operator whose kernel is the Green's function \(G(s,\tau)\), defined by
	\begin{equation}
		\left(-\frac{\mathrm{d}^{2}}{\mathrm{d}\tau^{2}}\right)G(s,\tau)=\delta(s-\tau) \,,
	\end{equation}
	such that we have 
	\begin{equation}
		\left(-\frac{\mathrm{d}^{2}}{\mathrm{d}\tau^{2}}\right)^{-1}\left(\lambda_{1}\left(\tau\right)-\lambda_{2}\left(\tau\right)\right)^{2}=\int \mathrm{d}s\,G(s,\tau)\left(\lambda_{1}\left(s\right)-\lambda_{2}\left(s\right)\right)^{2}\,.
	\end{equation}
	
This implies that we now need to solve for the Green's function. Time-translation invariance gives \(G(s,\tau)=G(r)\) where \(r=\lvert s-\tau \rvert\). It is easier to solve for the Green's function in Fourier space to give
	\begin{equation}\label{Green_int1}
		G(r)=\int \frac{\mathrm{d}\omega}{2\pi} \,\frac{e^{i\omega r}}{\omega^{2}}\, .
	\end{equation}
	We can turn the Fourier transform into a Gaussian integral by using the Schwinger representation $\frac{1}{p^{2}}=\int_{0}^{\infty} \mathrm{d}\alpha\, e^{-\alpha p^{2}}$, so that we get
	\begin{equation}
		\begin{aligned}
			G(r)&= \int_{0}^{\infty}\mathrm{d}\alpha \int \frac{\mathrm{d}\omega}{2\pi}\, e^{-\alpha \omega^{2}+i\omega r}\\
			&=\int_{0}^{\infty}\mathrm{d}\alpha \int \frac{\mathrm{d}\omega}{2\pi}\,e^{-\alpha\left(\omega-\frac{ir}{2\alpha}\right)^{2}+\frac{r^{2}}{4\alpha}}\,.
		\end{aligned}
	\end{equation}
The integral over \(\omega\) is now a Gaussian integral and gives \(\frac{1}{2\pi}\sqrt{\frac{\pi}{\alpha}}\). We are left with the \(\alpha\) integral, and under the change of variables \(\alpha=\frac{r^{2}}{4u}\), \(\mathrm{d}\alpha=-\frac{r^{2}}{4u^{2}}\mathrm{d}u\), and on doing some algebra, we have
	\begin{equation}
		G(r)=\frac{r}{4\sqrt{\pi}}\int_{0}^{\infty}\mathrm{d}u\, u^{-1/2-1}e^{-u}\,.
	\end{equation}
The integral is the definition of the Gamma function \(\Gamma(-\frac{1}{2})\) and has the value of \(-2\sqrt{\pi}\). Finally, we obtain the Green's function
	\begin{equation}\label{GreenFunct_massless}
		G(s,\tau)=-\frac{1}{2}\lvert s-\tau \rvert\,.
	\end{equation}
One could have also solved 	\eqref{Green_int1} by deforming the integral onto the complex plane, and used Cauchy's theorem, to arrive at the same result.
	
Expanding the logarithm in terms of the Green's function and taking the trace yields the following contribution to the effective action
\begin{eqnarray}\label{eq:expansion}
			&&\mathrm{Tr}\left[\log \left(1+\frac{1}{l_{s}^{4}}\int \mathrm{d}s\,G(s,\tau)(\lambda_{1}(s)-\lambda_{2}(s))^{2}\right)\right]\\
			&&=\frac{1}{l_{s}^{4}}\int \mathrm{d}\tau\, G(\tau,\tau)(\lambda_{1}(\tau)-\lambda_{2}(\tau))^{2}\nonumber\\
   && \;\;\;\;\;\;\;\;
			-\frac{1}{2l_{s}^{8}}\int \mathrm{d}s\mathrm{d}\tau\,  G(\tau,s)(\lambda_{1}(s)-\lambda_{2}(s))^{2}G(s,\tau)(\lambda_{1}(\tau)-\lambda_{2}(\tau))^{2}+\dots \nonumber
\end{eqnarray}
Plugging in the expression for \(G(s,\tau)\) that we derived above, we see that the first term vanishes and the second term gives 
\begin{equation}
		-\frac{1}{8l_{s}^{8}}\int \mathrm{d}s\mathrm{d}\tau\, \vert s-\tau\rvert^{2}(\lambda_{1}(s)-\lambda_{2}(s))^{2}(\lambda_{1}(\tau)-\lambda_{2}(\tau))^{2}
\end{equation}
which is non-local in time. Moreover, the series on the RHS of \eqref{eq:expansion} cannot be truncated as we cannot assume the factor inside the logarithm  to be small at all times, \textit{i.e.} \(\frac{1}{2l_{s}^{4}}\int \mathrm{d}s |s-\tau|(\lambda_{1}(s)-\lambda_{2}(s))^{2} \ll 1\) for all \(\tau \in (-\infty,\infty)\). Please see Appendix C for a more elaborate description of the time non-locality evident from this expression.
	
Let us now get back to the first term in the last line of \eqref{eq:trlog}. This can be understood as a product of the eigenvalues of the differential operator. In order to express the determinant as a product of all eigenvalues, we choose to work in periodic Euclidean time with period \(\beta\), and imposing periodic boundary conditions for the eigenfunctions restricts the $k$ values to be quantized as
	\begin{equation}\label{eq:discrete_k}
		k_{n}=\frac{2\pi n}{\beta} \quad n \in \mathbb{Z}\,.
	\end{equation}
Hence, a possible interpretation is that we are considering the matrix model in a finite temperature state. Using this expression, we get
	\begin{equation}
		\det \left(-\frac{\mathrm{d}^{2}}{\mathrm{d}\tau^{2}}\right)=\prod_{n=1}^{\infty}\left(\frac{2\pi n}{\beta}\right)^{2}\,,
	\end{equation}
which is a divergent constant term. We do not go into the merits of dropping this term as our bigger problem is the time non-local nature of the effective action described above.


\subsubsection{Adding a Mass Term}
	
If we add a term \(-m^{2}\mathrm{Tr}(Y^2) \) to the matrix Lagrangian, we find
	\begin{equation}
		S = \frac{1}{2l_{s}}\int d t \operatorname{Tr}\left[\left(D_t X\right)^2+\left(D_t Y\right)^2-\frac{1}{l_{s}^{4}}[X, Y]^2 -m^2 Y^2\right]\,.
	\end{equation}
There are two things to note here. Firstly, although we would ideally want to add only a mass term to the off-diagonal $Y_{12}$ terms alone so as to integrate them out, it is not possible to do so without also inducing a mass term to the diagonal elements of the $Y$ matrix (\textit{i.e.} to $\rho_i$). This is due to the nature of the invariant trace that the matrix action is endowed with. Secondly, by adding a mass term to the $Y$ matrix alone, we are of course explicitly breaking the symmetry between the two matrices in the original Lagrangian. It would probably be advisable to add masses to both the matrices to keep the symmetry between them. However, since the mass term here is simply a workaround to figure out if the general method works and we are treating this truly as a toy model, we keep the $X$ matrix massless for now. We discuss how we might be able to add mass terms in a covariant fashion in Sec-5.
	
Taking the trace and writing the action in terms of matrix elements explicitly, we have the Euclidean action
	\begin{eqnarray}
		S_{E} &=&\frac{1}{2l_{s}}\int \mathrm{d}\tau\, \Big(\dot{\lambda}_{1}^{2}+\dot{\lambda}_{2}^{2}+\dot{\rho}_{1}^{2}+\dot{\rho}_{2}^{2}+2\dot{Y}_{12}^{\ast} \dot{Y}_{12}+\frac{2}{l_{s}^{4}}\left(\lambda_{1} -\lambda_{2}\right)^{2} \lvert Y_{12}\rvert^{2}
   \nonumber\\
   & & \hspace{15mm} + m^2(\rho_1^2 + \rho_2^2 + 2 |Y_{12}|^2) \Big),
	\end{eqnarray}
	and \(\rho_1\), \(\rho_2\), \(Y_{12}\) are now massive.
On including this term, the effective action \eqref{eq:effective_action} takes the form
	\begin{eqnarray}
			e^{-S_{\textrm{eff}}\left[\lambda,\rho \right]} &:=& \int \mathcal{D} Y_{12}\, e^{-S_{E}\left[\lambda,\rho,Y_{12}\right]}\\
			&=& e^{-\frac{1}{2l_{s}}\int \mathrm{d}\tau\, \left(\dot{\lambda}_{1}^{2}+\dot{\lambda}_{2}^{2}+\dot{\rho}_{1}^{2}+\dot{\rho}_{2}^{2} +m^{2}(\rho_1^2 +\rho_2^2) \right)} \int \mathcal{D} Y_{12}\, e^{-\frac{1}{l_{s}}\int \mathrm{d}\tau\, \left( \dot{Y}_{12}^{\ast} \dot{Y}_{12}+\frac{1}{l_{s}^{4}}\left(\lambda_{1}-\lambda_{2}\right)^{2} \lvert Y_{12}\rvert^{2} + m^{2}|Y_{12}|^2\right)}\nonumber
	\end{eqnarray}
and the functional determinant now has an additional mass term. The effective action we obtain after integrating out \(Y_{12}\) is given by
	\begin{eqnarray}
		S_\textrm{eff}\left[\lambda,\rho\right]&=& \int \mathrm{d}\tau\, \frac{1}{2l_{s}}\left(\dot{\lambda}_{1}^{2} + \dot{\lambda}_{2}^{2}+ \dot{\rho}_{1}^{2} + \dot{\rho}_{2}^{2}+m^{2}(\rho_{1}^{2}+\rho_{2}^{2})\right)\nonumber\\
		&& -\frac{1}{2}\mathrm{Tr} \left[\log\left(\frac{1}{l_{s}} \left(-\frac{\mathrm{d}^{2}}{\mathrm{d}\tau^{2}} + m^{2} + \frac{1}{l_{s}^{4}}\left(\lambda_{1} - \lambda_{2}\right)^{2}\right)\right)\right]\,.
	\end{eqnarray}
Splitting the second term into two terms as before, we get 
\begin{eqnarray}\label{S_eff_mass}
	\mathrm{Tr}\left[\log\left(\frac{1}{l_{s}}\left(-\frac{\mathrm{d}^{2}}{\mathrm{d}\tau^{2}}+m^2+\frac{1}{l_{s}^{4}}\left(\lambda_{1}\left(\tau\right)-\lambda_{2}\left(\tau\right)\right)^{2}\right)\right)\right]=\mathrm{Tr}\left[\log\left(\frac{1}{l_{s}}\left(-\frac{\mathrm{d}^{2}}{\mathrm{d}\tau^{2}}+m^{2}\right)\right)\right]\nonumber\\ +\mathrm{Tr}\left[\log\left(1+\frac{1}{l_{s}^{4}}\left(-\frac{\mathrm{d}^{2}}{\mathrm{d}\tau^{2}}+m^2\right)^{-1}\left(\lambda_{1}\left(\tau\right)-\lambda_{2}\left(\tau\right)\right)^{2}\right)\right]\,.
\end{eqnarray}  
We find for the Green's function in this case by solving the equation
\begin{equation}
	\left(-\frac{\mathrm{d}^{2}}{\mathrm{d}\tau^{2}}+m^2\right)G(s,\tau)=\delta(s-\tau)\,.
\end{equation}
In Fourier space, we get 
\begin{equation}
	G(s,\tau)=\int \frac{\mathrm{d}\omega}{2\pi}\, \frac{e^{i\omega \lvert s- \tau \rvert}}{\omega^{2}+m^{2}}\,.
\end{equation}
To evaluate the integral, we push the contour up to wrap the upper branch cut on the imaginary axis starting at \(+im\), and we get
\begin{equation}\label{GreenFunct_massive}
		G(s,\tau)=\frac{1}{2m}e^{-m\lvert s-\tau \rvert}\,.
\end{equation}
	
Focusing on the second term on the right hand side of  \eqref{S_eff_mass}, for a large enough mass $m$ and $\lambda$'s sufficiently close, the Green's function will lead to an exponential damping of the integral inside the logarithm (see Appendix C for more about the physical interpretation of this). We can now expand the logarithm in a series and take the trace. Using the Green's function we found above, we have
	\begin{equation}\label{eq:expansion2}
		\begin{aligned}
			\mathrm{Tr}\left[\log \left(1+\frac{1}{l_{s}^{4}}\int \mathrm{d}s\,G(s,\tau)(\lambda_{1}(s)-\lambda_{2}(s))^{2}\right)\right]
			=\frac{1}{2ml_{s}^{4}}\int \mathrm{d}\tau (\lambda_{1}(\tau)-\lambda_{2}(\tau))^{2}\\
			-\frac{1}{4m^{2}l_{s}^{8}}\int \mathrm{d}\tau\mathrm{d}\tau^{\prime}\,\left[e^{-2m\lvert \tau- \tau^{\prime}\rvert}(\lambda_{1}(\tau)- \lambda_{2}(\tau))^{2}(\lambda_{1}(\tau^{\prime})-\lambda_{2}(\tau^{\prime}))^{2}\right]+\mathcal{O}(m^{-3})
		\end{aligned}
	\end{equation}
where we assumed \(\frac{1}{ml_{s}^{4}}\int \mathrm{d}s\,e^{-m|s-\tau|}(\lambda_{1}(s)-\lambda_{2}(s))^{2} \ll 1\) for all \(\tau \in (-\infty,\infty)\).
It is now safe to assume that the above series (\ref{eq:expansion2}) can be truncated since \(e^{-m|s-\tau|}\) is always less or equal to $1$ for positive values of $m$, and one simply needs to tune $m$ depending on the separation of  $(\lambda_{1}(\tau)-\lambda_{2}(\tau))^{2}$ to achieve this. Contrast this with the massless case where we cannot truncate the series (\ref{eq:expansion}) because the Green's function is of the form \(-\frac{1}{2}|s-\tau|\) and we cannot assume \(\frac{1}{2l_{s}^{4}}\int \mathrm{d}s\,|s-\tau|(\lambda_{1}(s)-\lambda_{2}(s))^{2} \ll 1\) for all values of \(\tau \in (-\infty,\infty)\). 

Applying our approximation to the RHS of the expression \eqref{eq:expansion2} above, we can truncate the series to leading order\footnote{To the leading order, our assumption translates into the condition \(\frac{1}{2ml_{s}^{4}}\int \mathrm{d}\tau\,(\lambda_{1}(\tau)-\lambda_{2}(\tau))^{2} \ll 1\), or equivalently \(\left[\frac{1}{2m}\int \mathrm{d}\tau\,(\lambda_{1}(\tau)-\lambda_{2}(\tau))^{2}\right]^{1/4} \ll l_{s}\).}  such that we arrive at the effective action with a potential term of the form:
\begin{equation}\label{eq:eff_action_massive}
		S_{\textrm{eff}}=\int \mathrm{d}\tau\, \left(\frac{1}{2l_{s}}\left(\dot{\lambda}_{1}^{2}+\dot{\lambda}_{2}^{2}+\dot{\rho}_{1}^{2}+\dot{\rho}_{2}^{2}+m^{2}(\rho_{1}^{2}+\rho_{2}^{2})\right)+\frac{1}{4ml_{s}^{4}}\left(\lambda_{1}-\lambda_{2}\right)^{2}\right).
\end{equation}
	
We need to clarify a small puzzle at this point. The Green's function obtained in \eqref{GreenFunct_massive} above does not reduce to that for the massless case in \eqref{GreenFunct_massless} in the limit of $m\rightarrow 0$. However, note that the Green's function is just a formal mathematical solution, and it needs to be integrated over to find the effective action or construct correlation functions from it. The effective action for the massless case indeed has the same divergence in it that appears when we take $m\rightarrow 0$ in \eqref{eq:eff_action_massive} above. 


\subsubsection{Regaining a cosmological constant term from the mass}

Going back to the first term on the RHS of \eqref{S_eff_mass}, we can rewrite the determinant as the product of all eigenvalues of the differential operator in the presence of the mass term, namely \(\left(-\frac{\mathrm{d}^{2}}{\mathrm{d}\tau^{2}}+m^{2}\right)\). The eigenfunctions of this differential operator are \(e^{ik\tau}\), and this gives the eigenvalue equation
\begin{equation}
	\left(-\frac{\mathrm{d}^{2}}{\mathrm{d}\tau^{2}}+m^{2}\right)e^{ik\tau}=\left(k^{2}+ m^{2}\right)e^{ik\tau}\,.
\end{equation}
Once again, choosing periodic boundary conditions and restricting the $k$ values as $k_{n}=\frac{2\pi n}{\beta},  n \in \mathbb{Z}$, we can rewrite the path integral as an infinite product:
\begin{equation}
	\begin{aligned}
		Z&\sim \prod_{n=-\infty}^{+\infty}\left[\left(\frac{2\pi n}{\beta}\right)^{2}+m^{2}\right]^{-1/2}\\
		&=\frac{1}{m}\prod_{n=1}^{\infty}\left[\left(\frac{2\pi n}{\beta}\right)^{2}+m^{2}\right]^{-1}\\
		&=\frac{1}{m}\prod_{n^{'}=1}^{\infty}\left(\frac{2\pi n^{'}}{\beta}\right)^{-2}\prod_{n=1}^{\infty}\left[1+\left(\frac{\beta m}{2\pi n}\right)\right]^{-1}\,.
	\end{aligned}
\end{equation}
The second of the two infinite products is convergent and has a well known result, 
\begin{equation}\label{eq:eigenvalue_eq}
	\prod_{n=1}^{\infty}\left[1+\left(\frac{\beta m}{2\pi n}\right)\right]=\frac{2}{\beta m}\sinh\left(\frac{\beta m}{2}\right)
\end{equation}
whereas the first infinite product is divergent but can be rendered finite through the technique of zeta function regularization \cite{Tong:susy}. The regularized end result is given by
\begin{equation}
	\prod_{n^{'}=1}^{\infty}\left(\frac{2\pi n^{'}}{\beta}\right)^{-2}=\frac{1}{\beta}\,.
\end{equation}
Putting the two pieces together, the path integral is evaluated to be
\begin{equation}
	Z=\frac{2}{\beta^2 m^2}\frac{1}{2\sinh\left(\beta m /2\right)}\,,
\end{equation}
which is nothing but a constant additive term to the effective action. It is well-known that putting in a mass term for the matrices results in an infrared cutoff which acts as a cosmological constant. Here, we have shown this result from a completely different perspective whereby integrating out the off-diagonal terms, after some regularization, results in a cosmological constant term to the overall action. However, we will drop this term from hereon and not keep it as a constant term in the action when going over to the collective field (in the emergent $(2+1)$-d) since it is an artefact of the mass term that we have added to one of the matrices. Moreover, such a constant term in the action does not affect the correlation functions.


\subsection{Generalization to arbitrary \texorpdfstring{$N$}{N}}
 In the above we assumed $N=2$ for simplicity, but the generalization to $N \times N$ matrices is straightforward. Analogous to \eqref{eq:effective_action}, the Euclidean action for 2 massless $N \times N$ matrices after taking the trace is given by
\begin{equation}
    S_{E}=\frac{1}{2l_{s}}\int \mathrm{d}\tau\, \left(\sum_{i}^{N}\dot{\lambda}_{i}^{2}+\sum_{i}^{N}\dot{\rho}_{i}^{2}+2\sum_{i<j, i=1}^{N}\dot{Y}_{ij}^{\ast} \dot{Y}_{ij}+\frac{2}{l_{s}^{4}}\sum_{i<j, i=1}^{N}\left(\lambda_{i}-\lambda_{j}\right)^{2} \lvert Y_{ij}\rvert^{2} \right) \,.
\end{equation}
The procedure for integrating out each \(Y_{ij}\) in the path integral is the same, resulting in the effective action
\begin{eqnarray}
    S_{\textrm{eff}}&=&\int \mathrm{d}\tau\, \left[\frac{1}{2l_{s}}\left(\sum_{i}^{N}\dot{\lambda}_{i}^{2}+\sum_{i}^{N}\dot{\rho}_{i}^{2}\right)\right]-\frac{1}{2}\sum_{i<j,i=1}^{N}\mathrm{Tr}\log\left(\frac{1}{l_s}\left(\frac{\rm d^2}{{\rm d}\tau^2}\right)\right)\nonumber\\
    && \;\;\;\;\;\;\;\;\;\;  -\frac{1}{2}\sum_{i<j,i=1}^{N}{\rm Tr}\log\left(1+\frac{1}{l_s^4}\int {\rm d}s\, G(s,\tau)\left(\lambda_{i}-\lambda_{j}\right)^{2}\right)\,.
\end{eqnarray}
The sum in the second term can be expressed explicitly in terms of N, which gives \(\frac{N(N-1)}{2}\) copies of the constant additive term that acts as a cosmological constant. After dropping this constant term, the time non-local effective action for the massless case can be formally written as
\begin{eqnarray}
    S_{\textrm{eff}}&=&\int \mathrm{d}\tau\, \left[\frac{1}{2l_{s}}\left(\sum_{i}^{N}\dot{\lambda}_{i}^{2}+\sum_{i}^{N}\dot{\rho}_{i}^{2}\right)\right] \nonumber\\
    & &\;\;\;\;\; \;\;\;\; -\frac{1}{2}\sum_{i<j,i=1}^{N}{\rm Tr}\log\left(1-\frac{1}{2l_s^4}\int {\rm d}s\, |s-\tau|\left(\lambda_{i}(s)-\lambda_{j}(s)\right)^{2}\right)\,.
\end{eqnarray}
If we add a mass term to the Y matrix, we argued that it allows us to expand the non-local logarithmic potential and truncate it to first order to obtain an effective action that is local in time. This would be analogous to \eqref{eq:eff_action_massive}, and we find
\begin{equation}
    S_{\textrm{eff}}=\int \mathrm{d}\tau\, \left(\frac{1}{2l_{s}}\left(\sum_{i}^{N}\dot{\lambda}_{i}^{2}+\sum_{i}^{N}\dot{\rho}_{i}^{2}+m^{2}\sum_i^N\rho_i^2\right)+\frac{1}{4ml_{s}^{4}}\sum_{i<j,i=1}^{N}\left(\lambda_{i}-\lambda_{j}\right)^{2}\right)\, .
\end{equation}

Although we will mainly work with the toy model involving the mass term for the $Y$ matrix in the next section, we remind the readers that for the two-matrix model analogous to the BFSS action, the effective action turns out to be non-local in time.


\section{The \texorpdfstring{$(2+1)$}{2+1}-dimensional collective field  action}

As was reviewed for the single matrix case in Appendix B, we can diagonalize the $X$ matrix using the $U(N)$ gauge symmetry. The Jacobian corresponding to the transformation \((X,Y) \rightarrow (\lambda,UYU^{\dagger})\) where \(\lambda\) is a diagonal matrix with non-degenerate eigenvalues $\lambda_{i}, (i=1,\ldots ,N)$, is given by the square of the Vandermonde determinant
	\begin{equation}
		\Delta(\lambda)=\prod_{1\leq i< j\leq N}(\lambda_{i}-\lambda_{j})\, .
	\end{equation}
To see the origin of this, note that the inner product of the wave functions must remain invariant under this $U(N)$ transformation, implying that 
\begin{eqnarray}
		&& \int {\rm d}X\,{\rm d}Y\, \Psi^{\ast}(X,Y)\Psi(X,Y)\nonumber\\
		&=& \int {\rm d} X_{\rm gf} \,\, {\rm d}Y_{\rm gf}\,\,\Delta^{2}\, \Psi^{\ast}\left(F(X_{\rm gf}), F(Y_{\rm gf})\right) \Psi\left(F(X_{\rm gf}), F(Y_{\rm gf})\right)\nonumber \\
        &=& \int {\rm d}\lambda \,\,{\rm d} Y_{\rm gf}\,\,\Psi^{\prime\ast}\left(\lambda, Y_{\rm gf}\right) \Psi^{\prime}\left(\lambda,Y_{\rm gf}\right)\,,
\end{eqnarray}
where \(F(X_{\rm gf})=X\) and \(F(Y_{\rm gf})=Y\). We see that the wave functions in the \((X_{\rm gf}, Y_{\rm gf})\) basis picks up the factor \(\Delta \)
\begin{equation}
      \Psi^{\prime}(X_{\rm gf}, Y_{\rm gf}) = \Delta(\lambda_i)\Psi(X,Y)\,.
\end{equation}
We have kept the definition of $X_{\rm gf} =: \lambda$ and $Y_{\rm gf}$ the same as in \eqref{diag_X} and \eqref{diag_X_basis}, respectively. Since the wavefunction of the bosonic matrices \(\Psi(X,Y)\) is totally symmetric, and  \(\Delta(\lambda_i)\) is anti-symmetric under \(\lambda_i \leftrightarrow \lambda_j\), the wavefunction govering the variables \((\lambda_i;\rho_i)\) is anti-symmetric in the \(\lambda\) variables and can be expressed as 
\begin{equation}
    \Psi(\lambda_1,\cdots,\lambda_N;\rho_1,\cdots,\rho_N)=\Delta(\lambda)\Phi(\lambda_1,\cdots,\lambda_N;\rho_1,\cdots,\rho_N)
\end{equation}
where \(\Phi\) is a totally-symmetric wavefunction. The anti-symmetric factor \(\Delta(\lambda)\) can be absorbed into the Hamiltonian such that the redefined Hamiltonian acts on the symmetric wavefunction \(\Phi\), under the redefinition
\begin{equation}\label{eq:similarity}
    H_{\text{eff}}=\frac{1}{\Delta} H \Delta.
\end{equation}

Recall that the goal of the previous section was to integrate out the off-diagonal pieces so as to write the action in terms of the diagonal entries of the two matrices alone, labelling the diagonal elements of the $Y$ matrix by $\rho_i$. It is important to emphasize that we are \textbf{not} diagonalizing $Y$, as indeed we cannot, since we have already diagonalized the $X$ matrix. However, we have integrated out the elements of the second matrix to find an effective action for $(\lambda_i,\rho_i)$ and their conjugate momenta. The difference in the footing of $X$ and $Y$ is now evident from the fact that the anti-symmetric Vandermonde determinant is only a function of $\lambda_i$ alone which must be  absorbed into the wavefunction of the D0-branes, \textit{i.e.} \(\Psi^{\prime}(\lambda_{1},\dots,\lambda_{N};\rho_{1},\dots,\rho_{N})= \Delta(\lambda_{1},\dots,\lambda_{N}) \Psi(X; Y)\). 
Consequently, this implies that the wavefunction is anti-symmetric under exchanges of $\lambda_i \leftrightarrow \lambda_j$ while it is symmetric under $\rho_i \leftrightarrow \rho_j$. Naturally, there is no dependence on the off-diagonal elements since we have integrated them out.

The above argument shows why much of this will be generalizable to the case of three or more matrices. Since we are only allowed to diagonalize one of the matrices, the Vandermonde determinant picks up the anti-symmetrization factor corresponding to that particular matrix alone. In other words, the diagonal elements of the other matrices will automatically be bosonic-valued. In our particular model of two matrices, we end up with a collective field theory emerging from $N$ fermions and $N$ bosons. However, the effect of integrating out the off-diagonal strings is that this is now a theory of {\bf $N$ interacting fermions}. In the general case of $d$ matrices, since we shall always only be able to diagonalize one of the matrices, the eigenvalues corresponding to it will constitute the $N$ interacting fermions.

Starting from the Euclidean effective action in \eqref{eq:eff_action_massive}, we go to the Hamiltonian picture and apply the collective field method. The Hamiltonian is given by (where we go back to explicitly choosing $N=2$ for brevity, but will quote the result for arbitrary $N$ at the end):
	\begin{equation}
		H=-\frac{l_{s}}{2}\left(\frac{1}{\Delta}\partial^2_{\lambda_1}\Delta+\frac{1}{\Delta}\partial^2_{\lambda_2}\Delta+\partial^{2}_{\rho_{1}}+\partial^{2}_{\rho_{2}}\right)+\frac{1}{2l_s}m^2(\rho_{1}^{2}+\rho_{2}^{2})+\frac{1}{4ml_{s}^{4}}(\lambda_{1}-\lambda_{2})^{2}
	\end{equation}
where \(\Delta = (\lambda_{1}-\lambda_{2})\) for $N=2$. We have applied the similarity transformation according to (\ref{eq:similarity}), and it only modifies the kinetic terms for the \(\lambda\)'s to \(\sum_{i}\partial^{2}_{i}=\sum_{i} \Delta \partial^{2}_{i}\frac{1}{\Delta}\).
	
The collective field method makes a change of variables from a set of discrete variables \((\lambda_{1},\dots,\lambda_{N};\, \rho_{1},\dots,\rho_{N})\) to a continuous field \(\phi(x,y)\). We have denoted our conventions for doing this change of variables in Appendices A and B, where we have reviewed the general formalism of the collective field theory and how it was applied to the single matrix case (corresponding to $N$ free fermions), respectively.  We define the collective field as a sum of Delta functions in a two-dimensional space of eigenvalue distributions: 
	\begin{equation}
		\phi(x,y)=\sum_{i=1}^{N}\delta(x-\lambda_{i})\delta(y-\rho_{i})\,,
	\end{equation}
which leads to the modified density constraint 
	\begin{equation}\label{eq:constraint1}
		\int \mathrm{d}x\mathrm{d}y\, \phi(x,y)=N\,.
	\end{equation}
This represents $N$ $D0$-branes located in 2 dimensional space at the coordinates given by the eigenvalues $\lambda_i$ and $\rho_i$.
As is standard, we make a point canonical transformation to the kinetic term\footnote{The eignevalues of the two matrices run from $-L/2$ to $L/2$, and hence this length scale appears on the RHS of \eqref{eq:kinetic_pt_canonical}. We would ultimately enforce the double scaling limit which is given by the usual $N\rightarrow \infty, L\rightarrow \infty$ while holding $N/L = {\rm const}.$}
	\begin{eqnarray}\label{eq:kinetic_pt_canonical}
		-\frac{1}{2}\sum_{i}\partial_{i}^{2}&=&\frac{1}{2 L^2}\int \mathrm{d}x\mathrm{d}y\, \partial^{2}_{i}\phi(x,y)\frac{\delta}{\delta\phi(x,y)}\nonumber\\
  & &-\frac{1}{2 L^4}\int \mathrm{d}x\mathrm{d}y\mathrm{d}x^{\prime}\mathrm{d}y^{\prime}\, \partial_{i}\phi(x,y)\partial_{i}\phi(x^{\prime},y^{\prime})\frac{\delta}{\delta\phi(x,y)}\frac{\delta}{\delta\phi(x^{\prime},y^{\prime})}\, .
	\end{eqnarray}
	From the above, we can define:
	\begin{equation}
		\begin{aligned}
			\omega(x,y)&=-\sum_{i}\partial_{i}^2\phi(x,y)\,,\\			\Omega(x,y,x^{\prime},y^{\prime})&=\sum_{i}\partial_{i}\phi(x,y)\partial_{i}\phi(x^{\prime},y^{\prime})\, .
		\end{aligned}
	\end{equation}
However, unlike in the single matrix model, in this case the \(\omega\)'s coming from the \(\lambda_{i}\) kinetic terms would be modified due to the Vandermonde determinant, but the ones from the \(\rho_{i}\) kinetic terms remain unchanged. The \(\Omega\)'s for all the terms remain the same due to cancellations between various terms. Explicitly, we have 
	\begin{equation}
		\begin{aligned}
			\omega_{\lambda}(x,y)&=-\sum_{i}\frac{1}{\Delta}\partial_{\lambda_i}^2\Delta\phi(x,y)\,,\\
			\Omega_{\lambda}\,,(x,y,x^{\prime},y^{\prime})&=\sum_{i}\partial_{\lambda_i}\phi(x,y)\partial_{\lambda_i}\phi(x^{\prime},y^{\prime})\,,
		\end{aligned}
	\end{equation}
	and
	\begin{equation}
		\begin{aligned}
			\omega_{\rho}(x,y)&=-\sum_{i}\partial_{\rho_i}^{2}\phi(x,y)\,,\\
			\Omega_{\rho}(x,y,x^{\prime},y^{\prime})&= \sum_{i}\partial_{\rho_i}\phi(x,y)\partial_{\rho_i}\phi(x^{\prime},y^{\prime})\,.
		\end{aligned}
	\end{equation}

To take into account the Jacobian that arises from the change of variables (point canonical transformation) to the collective field, we follow the general formalism (summarized in Appendix A) and apply another similarity transformation \(J^{1/2}HJ^{-1/2}\) where J is the Jacobian of the transformation to the collective field. Hence, the operator \(\frac{\delta}{\delta\phi(x,y)}\) transforms as
\begin{equation}
    \frac{\delta}{\delta\phi(x,y)}\rightarrow J^{1/2}\frac{\delta}{\delta\phi(x,y)} J^{-1/2}=\frac{\delta}{\delta\phi(x,y)}-\frac{1}{2}\frac{\delta\ln J}{\delta\phi(x,y)}\,.
\end{equation}
Substituting this into (\ref{eq:kinetic_pt_canonical}) and denoting the conjugate momentum as \(\pi(x,y)=\frac{1}{i}\frac{\delta}{\delta\phi(x,y)}\), the Hamiltonian becomes
\begin{eqnarray}
    H &=& \frac{1}{2}\left[\frac{i}{L}\int\mathrm{d}x\,\mathrm{d}y\,\left(\omega_\lambda(x,y)+\omega_{\rho}(x,y)\right)\left(\pi(x,y)+i\frac{1}{2}\frac{\delta \ln J}{\delta\phi(x,y)}\right)\right.\nonumber\\
    & & \left. \vspace{3mm}+ \frac{1}{L^{3}}\int \mathrm{d}x\,\mathrm{d}y\,\mathrm{d}x^{\prime}\,\mathrm{d}y^{\prime}\,\left(\Omega_{\lambda}(x,y,x^\prime,y^\prime)+\Omega_{\rho}(x,y,x^\prime,y^\prime)\right)\left(\pi(x,y)+i\frac{1}{2}\frac{\delta \ln J}{\delta\phi(x,y)}\right) \times \right. \nonumber\\ 
    & & \hspace{10mm} \left. \times \left(\pi(x^\prime,y^\prime)+i\frac{1}{2}\frac{\delta \ln J}{\delta\phi(x^\prime,y^\prime)}\right)\right]+V\,.
\end{eqnarray}
Similar to (\ref{eq:lnJ}) in Appendix A, instead of directly computing the Jacobian, the shortcut is to use the Hermiticity condition of the Hamiltonian to express the Jacobian in terms of \(\omega\)'s and \(\Omega\)'s as
\begin{equation}
    -\frac{\delta\ln J}{\delta\phi(x^\prime,y^\prime)}=\frac{1}{L^2}\int\mathrm{d}x\,\mathrm{d}y\,\Omega^{-1}(x,y,x^\prime,y^\prime)\left(\omega(x,y)+\frac{1}{L^{2}}\int\mathrm{d}x^{\prime\prime}\,\mathrm{d}y^{\prime\prime}\,\frac{\delta\Omega(x,y,x^{\prime\prime},y^{\prime\prime})}{\phi(x^{\prime\prime},y^{\prime\prime})}\right).
\end{equation}
Using the above, the leading order contribution to the Hamiltonian in the large N limit is
\begin{equation}\label{eq:eff_Hamiltonian}
		\begin{aligned}
			H_\textrm{eff}=&\frac{l_s}{2L^4}\int \mathrm{d}x\,\mathrm{d}y\, \pi(x,y)(\omega_{\lambda}(x,y)+\omega_{\rho}(x,y))\pi(x,y)\\+&\frac{l_s}{8 L^4}\int\mathrm{d}x\,\mathrm{d}y\,\mathrm{d}x^{\prime}\,\mathrm{d}y^{\prime}\,\left(\omega_{\lambda}(x,y)\Omega_{\lambda}^{-1}(x,y,x^{\prime},y^{\prime})\omega_{\lambda}(x,y)+\omega_{\rho}(x,y)\Omega_{\rho}^{-1}(x,y,x^{\prime},y^{\prime})\omega_{\rho}(x,y)\right)\\+&\Tilde{V}[\phi]\,,
		\end{aligned}
	\end{equation}
 similarly to (\ref{eq:leading_H}) in Appendix A. \(\Tilde{V}[\phi]\) is the potential in terms of the collective field, and is expressed as
\begin{equation}
    \Tilde{V}[\phi]=\frac{1}{8ml_{s}^{4}}\int\mathrm{d}x\,\mathrm{d}y\,\mathrm{d}x^{\prime}\,\mathrm{d}y^{\prime}\,\phi(x,y)(x-x^{\prime})^{2}\phi(x,y)+\frac{m^{2}}{2l_s}\int \mathrm{d}x\,\mathrm{d}y\,y^2\phi(x,y)\,.
\end{equation}
Note that we have used the properties of the Delta functions to rewrite the potential term \(\sim (\lambda_1-\lambda_2)^2\) in terms of the collective field \(\phi(x,y)\). The second term of the effective Hamiltonian is the leading order interaction term, and is denoted by \(V_\textrm{coll}\). The effective Hamiltonian is symbolically expressed as
    \begin{equation}
        H_{\textrm{eff}}=K+V_{\textrm{coll}}+\Tilde{V}
    \end{equation}
	where K is the kinetic term.
	In position space, the \(\omega(x,y;\phi)\)'s and \(\Omega(x,y,x^{\prime},y^{\prime};\phi)\)'s can be evaluated to give
	\begin{equation}
		\begin{aligned}
			\omega_{\lambda}(x,y)&=\frac{2}{L^2}\partial_{x}\left(\phi(x,y)\,\mathcal{P}\int\mathrm{d}x^\prime\,\frac{\int\mathrm{d}y^\prime\,\phi(x^\prime,y^\prime)}{x-x^\prime}\right)\\
			\Omega_{\lambda}(x,y,x^{\prime},y^{\prime};\phi)&=\partial_{x}\,\partial_{x^{\prime}}\left(\delta(x-x^{\prime})\delta(y-y^{\prime})\phi(x,y)\right)\,,\\
			\omega_{\rho}(x,y;\phi)&=\partial_{y}^{2}\phi(x,y)\,,\\
			\Omega_{\rho}(x,y,x^{\prime},y^{\prime})&=\partial_{y}\,\partial_{y^{\prime}}\left(\delta(x-x^{\prime})\delta(y-y^{\prime})\phi(x,y)\right)\,,\\
		\end{aligned} 
	\end{equation}
	where the difference between the expressions for \(\omega_{\lambda}\) and \(\omega_{\rho}\) comes from Vandermonde determinant associated with the \(\lambda\)'s.
	
We evaluate the effective potential to find
	\begin{equation}
		\begin{aligned}
			V_{\textrm{eff}}&=V_{\textrm{coll}}+\Tilde{V}\\&=\frac{l_s}{2}\int \mathrm{d}x\,\mathrm{d}y\,\phi(x,y)\left[\mathcal{P}\int\mathrm{d}x^{\prime}\,\frac{\int\mathrm{d}y^{\prime}\phi(x^\prime,y^\prime)}{x-x^{\prime}}\right]^{2}+\frac{l_s}{8}\int \mathrm{d}x\,\mathrm{d}y\,\frac{\left(\partial_{y}\phi(x,y)\right)^2}{\phi(x,y)}\\&+\frac{1}{8ml_s^4}\int \mathrm{d}x\,\mathrm{d}y\,\mathrm{d}x^{\prime}\,\mathrm{d}y^{\prime}\, \phi(x,y)(x-x^{\prime})^{2}\phi(x^{\prime},y^{\prime})+\frac{m^{2}}{2l_{s}}\int \mathrm{d}x\,\mathrm{d}y\,y^{2}\phi(x,y)\,.
		\end{aligned}
	\end{equation}
	The first term can be further evaluated by performing the integrals over the primed variables to give a \(\phi^3(x,y)\) term. We use the formula \cite{Sakita:1985exh}
	\begin{equation}
		\int \mathrm{d}x\, \phi(x)\left[\mathcal{P}\int \mathrm{d}x^{\prime}\, \frac{\phi(x^{\prime})}{x-x^{\prime}}\right]^{2}=\frac{\pi^{2}}{3}\int \mathrm{d}x\, \phi^3(x)
	\end{equation}
	to get \(\frac{l_{s}\pi^2}{6}\int \mathrm{d}x\,\mathrm{d}y\, \phi^3(x,y)\).
	The kinetic term (first term in (\ref{eq:eff_Hamiltonian})) is found to be
	\begin{equation}
		K=\frac{l_s}{2 L^2}\int \mathrm{d}x\,\mathrm{d}y\,\left(\partial_{x}\pi(x,y)\phi(x,y)\partial_{x}\pi(x,y)+\partial_{y}\pi(x,y)\phi(x,y)\partial_{y}\pi(x,y)\right).
	\end{equation}
	Bringing everything together, we have the Hamiltonian for the (2+1)-d collective field
	\begin{eqnarray}\label{Final_result}
			H&=&\frac{l_s}{2 L^2}\int \mathrm{d}x\,\mathrm{d}y\,\left[\partial_{x}\pi(x,y)\phi(x,y)\partial_{x}\pi(x,y)+\partial_{y}\pi(x,y)\phi(x,y)\partial_{y}\pi(x,y)\right]\nonumber\\
           & & +\frac{l_s\pi^2}{6 L^2}\int \mathrm{d}x\,\mathrm{d}y\, \phi^3(x,y)+\frac{l_s}{8}\int \mathrm{d}x\,\mathrm{d}y\,\frac{\left(\partial_{y}\phi(x,y)\right)^2}{\phi(x,y)}\nonumber\\
           & &+\frac{1}{8ml_s^4}\int \mathrm{d}x\,\mathrm{d}y\,\mathrm{d}x^{\prime}\,\mathrm{d}y^{\prime}\, \phi(x,y)(x-x^{\prime})^{2}\phi(x^{\prime},y^{\prime})+\frac{m^{2}}{2l_s}\int \mathrm{d}x\,\mathrm{d}y\,y^{2}\phi(x,y)\nonumber\\
   & & -\nu\left[\int \mathrm{d}x\,\mathrm{d}y\,\phi(x,y)-N\right]\,,
	\end{eqnarray}
 where \(\nu\) is a Lagrange multiplier that imposes the constraint in (\ref{eq:constraint1}).

We note that we can recover the single matrix collective field effective Hamiltonian (\ref{eq:single_matrix_H}) (see Appendix B) from our two-matrix Hamiltonian considered above. If we eliminate the $y$ direction and replace the two-matrix interaction term (which is trivially zero in the single matrix model) with an arbitrary potential $V(M)$, where $M$ denotes the single Hermitian  matrix. The potential for the single matrix model can always be rewritten as \(\int \mathrm{d}x\, V(x)\phi(x)\) using the collective field \(\phi(x)=\sum_i \delta(x-\lambda_i)\) since a single matrix can always be diagonalized and be expressed only in terms of its eigenvalues, such that then we obtain the effective Hamiltonian for the single matrix model:
\begin{equation}
    H = \int\mathrm{d}x\,\left(\frac{l_s}{2 L^2} \partial_{x}\pi(x)\phi(x)\partial_{x}\pi(x)+ \frac{l_s\pi^2}{6}\phi^3(x)+V(x)\phi(x)\right)-\nu\left(\int \mathrm{d}x\,\phi(x)-N\right)\,,
\end{equation}
which agrees with (\ref{eq:single_matrix_H}) if we set \(l_s=1\) and $\nu =\mu_F$. This tells us that the new potential term, which has convolutions in the $x$-direction (the first term on the third line in \eqref{Final_result}), is due to the BFSS like quartic potential $[X,Y]^2$ and is naturally absent in the single matrix case. Clearly, this term has some emergent spatial non-locality in the $x$-direction.

We can also formally write down the collective field Hamiltonian for the massless, time non-local case. The time non-local potential can be formally expressed in terms of the collective field as
\begin{eqnarray}\label{Final_result1}
        && \mathrm{Tr}\left[\log \left(1+\frac{1}{l_{s}^{4}}\int \mathrm{d}s\,G(s,\tau)(\lambda_{1}(s)-\lambda_{2}(s))^{2}\right)\right]\\
        &=&\mathrm{Tr}\left[\log \left(1-\frac{1}{2l_s^4}\int \mathrm{d}s\, |s-\tau|(\lambda_1(s)-\lambda_2(s))^2\right)\right]\nonumber\\
        &=&\mathrm{Tr}\left[\log \left(1-\frac{1}{2l_s^4}\int \mathrm{d}s\,\mathrm{d}x\,\mathrm{d}y\,\mathrm{d}x^{\prime}\,\mathrm{d}y^{\prime}\, |s-\tau|\phi(x,y)\left(x(s)-x^{\prime}(s)\right)^2\phi(x^{\prime},y^{\prime})\right)\right] \,.\nonumber
\end{eqnarray}
The kinetic terms remain the same, and thus the effective potential for the collective field can be read from the term above. However, it is not possible to write a closed-form expression of this when a truncation of the formal expansion of the logarithm is not available. Thus, we leave dealing with this time non-local potential for future work.

\section{Conclusions}
The collective field for single matrix quantum mechanics is a description of the eigenvalue distribution, fluctuations around the classical vacuum of which gives the spectrum of a massless scalar field in $(1+1)$-d with position-dependent couplings. Alternatively, it can also be thought of a theory of $N$ non-interacting fermions. In this work, we have generalized this to the case a of two-matrix action which has the form of (the bosonic part of) the BFSS model. Our goal was to understand how two spatial dimensions emerge in this model in the large $N$ limit. In order to take advantage of the collective field formalism, we gauge-fixed the action first and then integrated out the off-diagonal elements, before ending up with a model of {\bf $N$ interacting fermions and $N$ bosons}. It is at this point that we invoked the collective field formalism.

One of our findings was that it is important to add a small mass term to at least one of the matrices in order to have a time-local effective action. Even in this case, due to the interactions between the different eigenvalues, the resulting collective field action can clearly be seen to have a term that have convolutions in space (see the first term on the third line of \eqref{Final_result}). Our next goal would be to find the vacuum of this Hamiltonian and thereby the classical configuration of $\phi(x,y)$ that minimizes the action. Only then would we be able to compute fluctuations around this potential, whereby the dispersion relation of the fluctuation field will tell us about the metric of this emergent $(2+1)$-d spacetime. This is a nontrivial task since finding the minima of a Hamiltonian with terms that are non-local in space does not have any unique prescription, something that will be tackled in our upcoming work. However, even without figuring out these details, it is clear how one can successfully use the collective field picture to understand how a higher-dimensional field theory can emerge or descend from the space of eigenvalues of the two matrices. 

A major limitation of our method is that for the BFSS-like model, the integrating out leads to a time non-local effective action for the diagonal elements of the matrices even before transforming to the collective field. Adding a small mass term, as we have done in this work, is not a satisfactory resolution as we end up spoiling the relationship of these models with string theory. One way around this problem would be to consider a BMN-type toy model of two matrices \cite{Berenstein_2002}, whose classical solutions are known to be given by the fuzzy sphere \cite{Frenkel:2023aft}. One can resort to solving the collective field for this model, after gauge-fixing it first, and explore the space of fluctuations in the density of eigenvalues in such a model. As our expression \eqref{Final_result1} indicates, there is no general obstruction in applying our method to compute the collective field Hamiltonian in this case; only a finite truncation of the effective potential cannot be found rendering the whole expression time non-local (and therefore, necessarily non-Markovian). This type of non-local effects are perhaps expected to arise in string theory \cite{Arvanitakis:2021ecw} and we also leave their resolution for future work.

Recently, there has been a considerable amount of work on understanding how matrices encode geometrical information by focusing on a gauge-invariant notion of target space entanglement in matrix models \cite{Hampapura:2020hfg, Das:2020xoa,Frenkel:2023aft,Fliss:2024don}. They share many of the technical simplifications used by us such as gauge-fixing the matrices first as well as looking at simpler toy models involving a few matrices. Another future direction for us would be to study the entanglement entropy of these multi-matrix models starting from the collective field picture. Although this has been done in the single matrix context \cite{Das:1995vj,Das:2022nxo, Hartnoll:2015fca, Mandal:2024xqn}, extensions to the multi-matrix case has only been attempted at the formal level. Finding how the entanglement entropy scales with the emergent spatial dimensions, and if indeed it is finite for the collective field QFT, would be necessary ingredients to show the convergence of these different approaches. 

Before concluding, note that interest in matrix models have rejuvenated in recent years for their role in understanding an emergent notion of gravity \cite{Hartnoll:2024csr,Komatsu:2024bop, Komatsu:2024vnb, Hattori:2024btt, Manta:2024vol, Battista:2023glw, Klinkhamer:2022frp, Steinacker:2023myp} and  cosmology \cite{Brahma_2022met, Brahma_2022cos, Laliberte:2023bai}. In understanding how a large $(3+1)$-d spacetime emerges in this formalism \cite{Hirasawa:2023lpb, Nishimura:2022alt, nishimura2020newperspectivesemergence31d, hirasawa2021newphaselorentziantype, hirasawa2024effectssusyemergentspacetime, Nishimura_2002},  of particular interest is the the origin of $SO(9) \rightarrow SO(3) \times SO(6)$ symmetry breaking which has been recently studied in the IKKT matrix model \cite{brandenberger2024originso9rightarrowso3, Laliberte:2024iof}, going beyond numerical simulations \cite{Anagnostopoulos:2022dak, Anagnostopoulos:2020xai, Kim:2011cr, Ambj_rn_2000}, and in the BFSS model \cite{Brahma:2022ifx}. Another interesting future research direction would be to explore such symmetry breaking patterns using the collective field theory method explored in this paper.  

\section*{Acknowledgments}
SB is supported in part by the Higgs Fellowship and by the STFC Consolidated Grant ``Particle Physics at the Higgs Centre''. The research at McGill is supported in part by funds from NSERC and from the Canada Research Chair program. JP is supported in part by the Walter C. Sumner Memorial Fellowship. SB thanks Sumit Das for comments on an earlier version of this draft. JP wishes to thank Jo$\tilde{\text{a}}$o Penedones, Antoine Vuignier, and Xiang Zhao for useful discussions.


\section*{App A: General formalism of Collective field Theory}\label{appendix:app_A}
We will review the general technique \cite{Jevicki:1979mb,Jevicki:1980zg,Sakita:1985exh} of collective field theory which applies to the single matrix model. The main idea is a change of variables from the original set of finite degrees of freedom \(\left\{q_{i}\right\}_{i=1}^{N}\) to a new set of scalar field variable \(\phi\left(x,\left\{q_i\right\}\right)\) that has infinitely many degrees of freedom, taking into account of the Jacobian and the constraint that arise from the change of variables. 

Consider a general N-body Hamiltonian 
\begin{equation}
    H=-\frac{1}{2}\sum_{i}^{N}\frac{\partial^2}{\partial {q^i}^2}+V(\left\{q\right\})
\end{equation}
 and make a change of variables from \(\left\{q_{i}\right\}\) to a collective vairable \(\phi\left(x,\left\{q_i\right\}\right)\). The change of variables to the kinetic term is done by applying the chain rule
\begin{equation}\label{eq:chain_rule}
    \begin{aligned}
        -\frac{1}{2}\sum_{i}^{N}\frac{\partial^2}{\partial {q^i}^2}&=-\frac{1}{2}\sum_i\,\int\mathrm{d}x\,\frac{\partial}{\partial q^i}\left(\frac{\partial\phi(x)}{\partial q^i}\frac{\delta}{\delta\phi(x)}\right)\\
        &=-\frac{1}{2}\left(\sum_i\,\int\mathrm{d}x\,\frac{\partial^2\phi(x)}{\partial {q^i}^2}\frac{\delta}{\delta\phi(x)}+\sum_{i,j}\,\int\mathrm{d}x\mathrm{d}x^{\prime}\,\frac{\partial\phi(x)}{\partial q^i}\frac{\partial\phi(x^\prime)}{\partial q^j}\frac{\delta^2}{\delta \phi(x)\delta \phi(x^\prime)}\right)\\
        &\equiv\frac{1}{2}\left(\int\mathrm{d}x\,\omega(x)\frac{\delta}{\delta\phi(x)}-\int\mathrm{d}x\mathrm{d}x^{\prime}\,\Omega(x,x^\prime)\frac{\delta^2}{\delta \phi(x)\delta \phi(x^\prime)}\right)
    \end{aligned}
\end{equation}
where \(\omega\) and \(\Omega\) are defined to be
\begin{equation}
    \begin{aligned}
        \omega(x)&=-\sum_i\,\frac{\partial^2\phi(x)}{\partial {q^i}^2}\\
        \Omega(x,x^\prime)&=\sum_{i,j}\,\frac{\partial\phi(x)}{\partial q^i}\frac{\partial\phi(x^\prime)}{\partial q^j}.\\
    \end{aligned}
\end{equation}
The collective field \(\phi(x)\) is subject to the constraint
\begin{equation}\label{eq:constraint}
    \int\mathrm{d}x\, \phi(x)=N.
\end{equation}
There is a Jacobian associated with the above change of variables that has to be taken into account. Under the change of variables, the inner products of the functions/functionals which the Hamiltonian acts on should be conserved
\begin{equation}
    \int [\mathrm{d}q]\,\psi^\ast(q)\psi(q)=\int J[\mathrm{d}\phi]\,\psi^\ast\left(F(\phi)\right)\psi\left(F(\phi)\right)=\int [\mathrm{d}\phi]\,\Psi^\ast(\phi)\Psi(\phi)
\end{equation}
where J is the Jacobian from the change of variables q to \(\phi\), and F is the inverse of the transformation, \(F(\phi)=q\). This gives the relation 
\begin{equation}
    \Psi(\phi)=J^{1/2}\psi(q)
\end{equation}
and the Hamiltonian that acts on \(\Psi(\phi)\) is obtained by the so-called similarity transformation
\begin{equation}\label{eq:similarity_transf}
    H_{\textrm{eff}}=J^{1/2}HJ^{-1/2}
\end{equation}
where H is the Hamiltonian that acts on \(\psi(q)\). Note that (\ref{eq:chain_rule}) is not Hermitian if we only make the change of variables without taking the Jacobian into account. 

The effect of (\ref{eq:similarity_transf}) on the functional derivative \(\frac{\delta}{\delta \phi(x)}\) is 
\begin{equation}
    J^{1/2}\frac{\delta}{\delta \phi(x)}J^{-1/2}=\frac{\delta}{\delta \phi(x)}-\frac{1}{2}\frac{\delta\ln J}{\delta\phi(x)}
\end{equation}
and substituting this into (\ref{eq:chain_rule}) gives

{\footnotesize
\begin{equation}\label{eq:similarity_K_lnJ}
    \begin{aligned}
        &\int\mathrm{d}x\,\omega(x)\left(\frac{\delta}{\delta \phi(x)}-\frac{1}{2}\frac{\delta\ln J}{\delta\phi(x)}\right)-\int\mathrm{d}x\,\mathrm{d}x^\prime\,\Omega(x,x^\prime)\left(\frac{\delta}{\delta \phi(x)}-\frac{1}{2}\frac{\delta\ln J}{\delta\phi(x)}\right)\left(\frac{\delta}{\delta \phi(x^\prime)}-\frac{1}{2}\frac{\delta\ln J}{\delta\phi(x^\prime)}\right)\\
        =&\frac{1}{2}\int\mathrm{d}x\,\left(\omega(x)+\int\mathrm{d}x^{\prime}\,\left(\frac{\delta\Omega(x,x^\prime)}{\delta\phi(x^\prime)}\right)+\int\mathrm{d}x^{\prime}\,\frac{\delta\ln J}{\delta\phi(x^{\prime})}\Omega(x,x^{\prime})\right)\frac{\delta}{\delta\phi(x)}\\&-\frac{1}{2}\int\mathrm{d}x\,\mathrm{d}x^{\prime}\,\left(\frac{\delta}{\delta\phi(x)}\Omega(x,x^\prime)\frac{\delta}{\delta\phi(x^\prime)}+\frac{1}{4}\frac{\delta\ln J}{\delta\phi(x)}\Omega(x,x^\prime)\frac{\delta\ln J}{\delta\phi(x^\prime)}\right)-\frac{1}{4}\int\mathrm{d}x\,\omega(x)\frac{\delta\ln J}{\delta\phi(x)}\\&+\frac{1}{4}\int\mathrm{d}x\,\mathrm{d}x^\prime\,\Omega(x,x^\prime)\left(\frac{\delta^2\ln J}{\delta\phi(x)\delta\phi(x^\prime)}\right)\\
    \end{aligned}
\end{equation}}
Since we would like to use a similarity transformation to write the Hamiltonian in a Hermitian form, we set the non-Hermitian terms which are in the second line of \eqref{eq:similarity_K_lnJ} to zero
\begin{equation}
    0=\int \mathrm{d}x\, \omega(x)\frac{\delta}{\delta \phi(x)}+\int \mathrm{d}x\,\mathrm{d}x^\prime\,\left(\frac{\delta\Omega(x,x^\prime)}{\delta \phi(x)}\right)\frac{\delta}{\delta\phi(x^\prime)}+\int \mathrm{d}x\,\mathrm{d}x^\prime\,\Omega(x,x^\prime)\left(\frac{\delta\ln J}{\delta\phi(x^\prime)}\right)\frac{\delta}{\delta\phi(x)}
\end{equation}
and arrive at the following expression for the Jacobian
\begin{equation}\label{eq:lnJ}
    -\frac{\delta\ln J}{\delta\phi(x^\prime)}=\int \mathrm{d}x\,\Omega^{-1}(x,x^\prime)\left(\omega(x)+\int\mathrm{d}x^{\prime \prime}\,\frac{\delta\Omega(x,x^{\prime \prime})}{\delta\phi(x^{\prime \prime})}\right)
\end{equation}
where the inverse of \(\Omega\) is defined as
\begin{equation}
    \int\mathrm{d}x^\prime\,\Omega^{-1}(x,x^\prime)\Omega(x^\prime,x^{\prime \prime})=\delta(x-x^{\prime \prime}).
\end{equation}
Substituting (\ref{eq:lnJ}) into the Hermitian terms in (\ref{eq:similarity_K_lnJ}) and combining terms gives the general expression for the full Hermitian Hamiltonian for the collective field
\begin{equation}
    \begin{aligned}
        H=&-\frac{1}{2}\int\mathrm{d}x\,\mathrm{d}x^\prime\,\frac{\delta}{\delta \phi(x)}\Omega(x,x^\prime)\frac{\delta}{\delta\phi(x^\prime)}\\&+\frac{1}{8}\int\mathrm{d}x\,\mathrm{d}x^\prime\,\left(\omega(x)+\int\mathrm{d}x^{\prime\prime}\,\frac{\delta\Omega(x,x^{\prime\prime})}{\delta\phi(x^{\prime\prime})}\right)\Omega^{-1}(x,x^\prime)\left(\omega(x^\prime)+\int\mathrm{d}x^{\prime\prime\prime}\,\frac{\delta\Omega(x^\prime,x^{\prime\prime\prime})}{\delta\phi(x^{\prime\prime\prime})}\right)\\
        &-\frac{1}{4}\int\mathrm{d}x\,\frac{\delta\omega(x)}{\delta\phi(x)}-\frac{1}{4}\int\mathrm{d}x\,\mathrm{d}x^\prime\,\frac{\delta^{2}\Omega(x,x^\prime)}{\delta\phi(x)\delta\phi(x^\prime)}+V\left[\phi\right].
    \end{aligned}
\end{equation}
Defining the conjugate momenta of the collective field \(\phi(x)\) as \(\pi(x)\equiv\frac{1}{i}\frac{\delta}{\delta\phi(x)}\), the first term is the kinetic term of the collective field \(\phi(x)\) while the rest of the terms are interaction potential terms. In the \(N\rightarrow\infty\) limit, rescaling the constraint (\ref{eq:constraint}) to make the N dependence explicit, the leading order Hamiltonian (terms with highest degree in N) is 
\begin{equation}\label{eq:leading_H}
    H=\frac{1}{2}\int\mathrm{d}x\,\mathrm{d}x^\prime\,\pi(x)\Omega(x,x^\prime)\pi(x^\prime)+\frac{1}{8}\int\mathrm{d}x\,\mathrm{d}x^\prime\,\omega(x)\Omega^{-1}(x,x^\prime)\omega(x^\prime)+V[\phi]
\end{equation}
where the first term \(\sim\pi\Omega\pi\) is the kinetic term and the second term \(\sim\omega\Omega^{-1}\omega\) is conventionally referred to as \(V_{\textrm{coll}}\).


\section*{App B: Collective field action for a single matrix model}
In \cite{Das:1990kaa}, Das and Jevicki derived the collective field action for the single matrix model and found it to have a physical interpretation of a scalar field in two dimensions. Here we review their derivation of the collective field Hamiltonian of the single matrix model for comparison to our two-matrix model results. In this appendix we use dimensionless variables for simplicity.

The idea is to change variables from the original matrix observables to a new set of scalar fields. The single matrix model has the quantum mechanical Hamiltonian:
\begin{equation}
H = -\frac{1}{2} \sum_{i,j=1}^N\frac{\partial^2}{\partial M_{ij}\partial M_{ji}}+\operatorname{Tr} \left( V(M)\right)\,,
\end{equation}
for some potential $V(M)$. We want to use the gauge-invariant set of variables ${\rm Tr} (M^k)$, of which only $N$ are linearly independent. However, in the $N\rightarrow \infty$ limit, they all become independent variables. Thus, one can introduce the collective field as a change of variables to
\begin{eqnarray}
    \phi_k = \operatorname{Tr}(e^{ikM})\,,
\end{eqnarray}
where $k$ is simply the `loop' index at this point. Under this transformation, the kinetic term can be rewritten as
\begin{equation}
\begin{aligned}
-\frac{1}{2} \sum_{i,j=1}^N\frac{\partial^2}{\partial M_{ij}\partial M_{ji}} =\frac{1}{2} \left(\int \mathrm{d}k\,\omega(k, \phi) \frac{\delta}{\delta \phi_k} - \int \mathrm{d}k\,\mathrm{d}k^\prime\,\Omega\left(k, k^{\prime} ; \phi\right) \frac{\delta}{\delta \phi_k} \frac{\delta}{\delta \phi_{k^\prime}}\right)\,,
\end{aligned}
\end{equation}
where
\begin{equation}
\begin{aligned}
\omega(k, \phi) &=-\sum_{i,j} \frac{\partial^2 \phi_k}{\partial M_{ij} \partial M_{ji}} \\
\Omega\left(k, k^{\prime} ; \phi\right) &=\sum_{i,j} \frac{\partial \phi_k}{\partial M_{ij}} \frac{\partial \phi_{k^\prime}}{\partial M_{ji}}\,.
\end{aligned}
\end{equation}
The above quantities come from applying the chain rule to the kinetic term. We can now simplify these quantities by using the formula
\begin{eqnarray}
    \frac{{\rm d} \left(e^M\right)_{kl}}{{\rm d} M_{ij}} = \int_0^1 {\rm d} a \left(e^{a M}\right)_{ik} \ \left(e^{(1-a) M}\right)_{jl} \ ,
\end{eqnarray}
which leads to 
\begin{equation}
\begin{aligned}
& \omega(k, \phi) = k^2 \int_0^1 d \alpha \phi_{\alpha k} \phi_{k(1-\alpha)} \\
& \Omega\left(k, k^{\prime} ; \phi\right) = k k^{\prime} \phi_{k+k^{\prime}} \,.
\end{aligned}
\end{equation}

For the single matrix case, we can diagonalize the matrix,  $M = U^{-1} \operatorname{diag}\left(\lambda_i\right) U$, such that the collective field can be written in terms of the eigenvalues, $\lambda_i$, as
\begin{equation}
\phi_k  =\sum_{i=1}^N e^{i k \lambda_i}\,.
\end{equation}
We note that the dynamics of the angular coordinates, denoted by $U$ above, do not enter into the singlet sector of the theory and will be ignored in the following\footnote{The momenta corresponding to $U$ do appear in the Hamiltonian but annihilate the singlet sector wavefunction, and can hence be dropped for our case.}. Before simplifying the above expressions in terms of the eigenvalues, let us also point out that in quantum mechanics, the inner products of wave functions should remain invariant under the above change of variables. This would mean rescaling the wave-functions by the appropriate power of the Jacobian of the transformation. To see this more explicitly, consider that the Jacobian corresponding to the change of variables is $J$. Then the invariance of the inner-products demand that $\int \mathcal{D}[M] \bar{\psi}(M) \psi(M) = \int \mathcal{D}[\phi] J \bar{\Psi}(\phi) \Psi(\phi)$. One can absorb the factor of the Jacobian in the wave-function $\Psi \rightarrow J^{1/2} \Psi$ which, however, implies that the momentum operator would also have to be appropriate rescaled by powers of the Jacobian, \textit{i.e.} $\frac{\delta}{\delta \phi} \rightarrow J^{1/2} \, \frac{\delta}{\delta \phi} \, J^{-1/2}$ which appears in the momentum term of the Hamiltonian. In the case of diagonalizing a matrix, the Jacobian is given by the (square of the) Vandermonde  determinant, as used in the following and in the main text. 

After diagonalisation, one can also take the Fourier transform of the above quantities to move to real space, where the description is given in terms of density of the eigenvalues, that is:
\begin{eqnarray}
    \phi_k & =&  \operatorname{Tr}\left(e^{ikM} \right) = \sum_{i=1}^N e^{i k \lambda_i}\,.
\end{eqnarray}
The position space collective field is
\begin{eqnarray}
\phi(x) & =&\int\frac{{\rm d}k}{2\pi} e^{-ikx} \operatorname{Tr}\left(e^{ikM} \right)= \operatorname{Tr}\left(\delta(x-M)\right) = \sum_{i=1}^N \delta\left(x-\lambda_i\right)\,.
\end{eqnarray}
The Fourier transform of $\omega(k,\phi)$ is given by
\begin{equation}
\omega(x, \phi)=2\partial_{x}\left(\phi(x)\,\mathcal{P}\int\mathrm{d}x^{\prime}\frac{\phi(x^\prime)}{x-x^{\prime}}\right) \,,
\end{equation}
while that for $\Omega(k,k^\prime; \phi)$ is:
\begin{equation}
\Omega(x, x', \phi)= \partial_x \partial_{x^\prime} \left( \phi(x) \ \delta(x-x')\right) \,.
\end{equation}

Using these expressions for $\omega(x,\phi)$ and $\Omega(x,x',\phi)$, the Hamiltonian can be written as,
\begin{equation}
\begin{aligned}
H_\phi= & \int {\rm d} x \left[\frac{1}{2} \partial_x \pi \phi \partial_x \pi+V(x)\phi(x)-\mu_F\left(\phi(x)-\frac{N}{V}\right)\right] \\
& +\frac{1}{2} \int {\rm d} x \ \phi(x)\left[\int {\rm d} x' \frac{\phi(x')}{(x-x')}\right]^2 .
\end{aligned}
\end{equation}
where $\mu_F$ is a Lagrange multiplier that enforces the density constraint.

The effective potential in the Hamiltonian is,
\begin{equation}
V_{\text {eff }}=\frac{1}{2} \int {\rm d} x \  \phi(x)\left[\int {\rm d} x' \frac{\phi(x')}{(x-x')}\right]^2-\int{\rm d} x \left(\mu_F-V(x)\right) \phi(x) .
\end{equation}
where the integral in the first term can be evaluated to arrive at,
\begin{equation}
V_{\text {eff }}=\int {\rm d} x \left[\frac{\pi^2}{6} \phi^3(x)-\left(\mu_F-V(x)\right) \phi(x)\right] 
\end{equation}
Together the Hamiltonian is\footnote{There are two more terms of lower order that we do not write down here.} 
\begin{equation}\label{eq:single_matrix_H}
    \begin{aligned}
        H_\phi= & \int {\rm d} x \left(\frac{1}{2} \partial_x \pi \phi \partial_x \pi+\frac{\pi^2}{6} \phi^3(x)+V(x) \phi(x)\right)-\mu_F\left(\int\mathrm{d}x\,\phi(x)-\frac{N}{V}\right).
    \end{aligned}
\end{equation}


\section*{App C: Truncating the effective action for local physics}
Let us understand in a bit more detail why adding the mass term is necessary to get (time)-local physics out of the matrix model when integrating out the off-diagonal strings. In order to integrate out a field to get an effective action, one needs to invert the kinetic operator of this field. If the momenta $k$ is restricted to be smaller than the mass $m$ of the field, then we can expand the propagator as 
\begin{eqnarray}
 \frac{1}{k^2+m^2} = \frac{1}{m^2} \left[1 - \left(\frac{k}{m}\right)^2 + \left(\frac{k}{m}\right)^4 + \ldots \right]\,.
\end{eqnarray}
However, this expansion breaks down if the field that is integrated out is a massless one.

Let us consider a quantum mechanical (Euclidean) action to clarify this further\footnote{We closely follow the treatment of \cite{David_Skinner} for this part.}:
\begin{eqnarray}
		S[\varphi,\sigma] = \int {\rm d}t \  \left\{\frac{1}{2}\dot{\varphi}^2 + \frac{1}{2}\dot{\sigma}^2 + \frac{1}{2} m^2 \sigma^2 + g\varphi^2\sigma^2\right\}\,.
\end{eqnarray}
	
After integrating out the heavy field, $\sigma$, the effective action is of the form
	\begin{eqnarray}
		S^{\rm eff}[\varphi] = \int {\rm d}t \ \left\{\frac{1}{2}\dot{\varphi}^2 +\frac{1}{2} \ln \det\left[\frac{{\rm d^2}}{{\rm d}t^2} - m^2 + 2g \varphi^2 \right]^{-1/2} \right\}\,,
	\end{eqnarray}
	where the effective potential can be expanded in a formal series as
	\begin{eqnarray}
		{\rm Tr} \ln \left[1+2g \left(\frac{{\rm d^2}}{{\rm d}t^2} - m^2\right)^{-1} \varphi^2\right] &=& 2g \int {\rm d}t  \ G(t,t) \varphi^2(t)\\
        & &+ 2 g^2 \int {\rm d}t \int {\rm d}s \ G(s,t) \varphi^2(t) G(t,s) \varphi^2(s) +\ldots\nonumber
	\end{eqnarray}
	Recall that the Green's function in this case is given by $G(t,s) =(2m)^{-1} \ e^{-m |t-s|}$. Focusing on the second term above,
	\begin{eqnarray}
		&& \mathcal{O}(g^2) \ \int {\rm d}t \int {\rm d}s \ G(s,t) \varphi^2(t) G(t,s) \varphi^2(s) \\
		&=& \mathcal{O}(g^2) \ \int {\rm d}t \int {\rm d}s\, G(t,s)^2 \varphi^2(t) \left\{\varphi^2(t) + 2 \varphi(t) \dot{\varphi} (t-s) + \left[\dot{\varphi}^2 + \varphi(t) \ddot{\varphi}(t)\right] (t-s)^2 + \ldots \right\}\,,\nonumber
	\end{eqnarray}
	where we have Taylor expanded the field $\varphi(s)$ around $s=t$. The reason behind doing this is to show that higher (time) derivatives of $\varphi$ are suppressed by additional factors of $m$ in the above expression. Although the explicit expressions are a bit messy, it is easy to see that the above expression is of the form
	\begin{eqnarray}
		&& \mathcal{O}(g^2) \ \int {\rm d}t \int {\rm d}s \ G(s,t) \varphi^2(t) G(t,s) \varphi^2(s) \nonumber\\
		&=& \int{\rm d}t \ \left[\frac{c_0}{m^3} \varphi^4(t) + \frac{c_2}{m^5} ({\rm 2 \ derivative \ terms}) + \frac{c_4}{m^7} ({\rm 4 \ derivative \ terms}) +\ldots \right] \,.
	\end{eqnarray}
The symmetric form of the Green's function (which comes from the form of the action) ensures that the odd derivatives do not contribute to the above. Thus, if the field varies slowly on time-scales of $m^{-1}$, then the above series can easily be truncated to any finite order to find a time-local expression. This immediately shows us that when $m\rightarrow 0$, this series can never be truncated and we have a truly time non-local effective theory on our hands. Note that such nonlocal effective field theories often pop up in String Theory in its various incarnations (see e.g. \cite{Arvanitakis:2021ecw}).

\printbibliography
\end{document}